\documentclass[draftclsnofoot,onecolumn]{IEEEtran}
\pdfoutput=1

\usepackage{amsmath,amsfonts,amssymb,enumerate,url} 
\usepackage{graphicx} 

\IEEEoverridecommandlockouts
 
\def\bp{\bar{p}}

\def\ie{{\em i.e.},~} 
\def\eg{{\em e.g.},~} 
\def\etc{{\em etc.}} 
\def\pmax{p_{\sf max}} 
\def\Utot{U_{\sf total}} 
\def\Umax{U_{\sf max}}

\newcommand{\pref}[1]{(\ref{#1})} 
 
\newcommand{\pd}[2]{\frac{\partial{#1}}{\partial{#2}}} 
 
\begin{document} 
 
\title{A Study of Non-Neutral Networks with Usage-based Prices}

\author{\small
	\IEEEauthorblockN{E. Altman \& P. Bernhard$^1$
		\thanks{$^1$ INRIA, 2004 Route des Lucioles, 06902 Sophia-Antipolis, France}
		\thanks{Email: \texttt{\{eitan.altman,pierre.bernhard\}@sophia.inria.fr}}
	}
	\and
	\IEEEauthorblockN{S. Caron \& G. Kesidis$^2$
		\thanks{$^2$ CS\&E and EE Depts, Pennsylvania State Univ., University Park, PA, 16802}
		\thanks{Email: \texttt{kesidis@engr.psu.edu}.}
		\thanks{The work by George Kesidis is supported in part by NSF CNS and
		Cisco Systems URP grants.}
	}
	\and
	\IEEEauthorblockN{J. Rojas-Mora$^3$
		\thanks{$^3$ Fac. of Econ. and Bus. Sci., Univ. of Barcelona, 08034 Barcelona, Spain}
		\thanks{Email: \texttt{jrojasmo7@alumnes.ub.edu}}
	}
	\and
	\IEEEauthorblockN{S. Wong$^4$
		\thanks{$^4$ Fac. of Law, Univ. of Coru\~na, 15071 A Coru\~na, Spain}
		\thanks{Email: \texttt{swong@udc.es}}
	}
\normalsize}

\maketitle

\begin{abstract} 
Hahn and Wallsten \cite{HW} wrote that network neutrality ``usually means 
that broadband service providers charge consumers only once for Internet 
access, do not favor one content provider over another, and do not charge 
content providers for sending information over broadband lines to end 
users.'' In this paper we study the implications of non-neutral behaviors 
under a simple model of linear demand-response to \emph{usage-based} 
prices. We take into account advertising revenues and consider both
cooperative and non-cooperative scenarios. In particular,
we model the impact of side-payments between service and content providers.
We also consider the effect of \emph{service discrimination} by access
providers, as well as an extension of our model to non-monopolistic content
providers.
\end{abstract}

\section{Introduction} 
\setcounter{footnote}{4}

Network neutrality is an approach to providing network access 
without unfair discrimination among applications, content or 
traffic sources.
Discrimination occurs when there are two applications, services or 
content providers that require the same network resources, but one 
is offered better quality of service (shorter delays, higher 
transmission capacity, \etc) than the other. 
How to define what is ``fair'' discrimination is still subject to 
controversy.\footnote{ 
	The recent decision on Comcast v. the FCC was expected to deal 
	with the subject of ``fair'' traffic discrimination, as the 
	FCC ordered Comcast to stop interfering with 
	subscribers' traffic generated by peer-to-peer networking applications. 
	The Court of Appeals for the District of 
	Columbia Circuit was asked to review this order by Comcast, 
	arguing not only on the necessity of managing scarce network resources, 
	but also on the non-existent jurisdiction of the FCC over network 
	management practices. The Court decided that the FCC did not have express 
	statutory authority over the subject, neither demonstrated that its 
	action was "reasonably ancillary to the [...] effective performance of 
	its statutorily mandated responsibilities". The FCC was deemed, then, 
	unable to sanction discriminatory practices on Internet's traffic 
	carried out by American ISPs, and the underlying case on the ``fairness'' of 
	their discriminatory practices was not even discussed.}
A preferential treatment of traffic is considered fair as long as 
the preference is left to the user.\footnote{ 
	Nonetheless, users are just one of many actors in the net neutrality 
	debate, which has been enliven [?] throughout the world by several 
	public consultations for new legislations on the subject. 
	The first one, proposed in the USA, was looking for the best means 
	of preserving a free and open Internet. The second one, 
	carried out in France, asks for different points of view over net 
	neutrality. A third one is intended to be presented by the EU during 
	summer 2010, looking for a balance on the parties concerned as users 
	are entitled to access the services they want, while ISPs and CPs 
	should have the right incentives and opportunities to keep investing, 
	competing and innovating. 
	See \cite{FCC_cons,fr_cons,eu_cons}.}
Internet Service Providers (ISPs) may have interest in traffic discrimination 
either for technological or economic purposes. Traffic congestion, especially 
due to high-volume peer-to-peer traffic, has been a central argument for ISPs 
against the enforcement of net neutrality principles. However, it seems many 
ISPs have blocked or throttled such traffic independently of congestion 
considerations. 
 
ISPs recently claimed that net neutrality acts as a disincentive for 
capacity expansion of their networks. In \cite{Cheng08}, the authors 
studied the validity of this argument and came to the conclusion that, 
under net neutrality, ISPs invest to reach a social optimal level, 
while they tend to under/over-invest when neutrality is dropped. 
In their setting, ISPs stand as winners while content providers (CPs) 
are left in a worse position, and users who pay the ISPs for preferential 
treatment are better off while other consumers have a significantly 
worse service. 
 
ISPs often justify charging CPs by quantifying the large amount of 
network resources ``big'' content providers use. On the other hand, 
the content a CP offers contributes to the demand for Internet access, 
and thus benefits the access providers. 
 
Many references advocate the use of the Shapley value as a fair way to share 
profits between the providers, see, \eg \cite{shap1,shap2}. One of the main 
benefits of this approach is that it yields Pareto optimality for all players, 
and requires in particular that CPs, many of whom receive third-party 
income such as advertising revenue from consumers' demand, help pay for the 
network access that makes this new income possible. 
 
In this paper, we focus on violations of the neutrality principles defined in 
\cite{HW} where broadband service providers 
\begin{itemize} 
\item charge consumers more than ``only once'' through usage-based pricing, and 
\item charge content providers through side-payments. 
\end{itemize} 
Within a simple game-theoretic model, we examine how regulated\footnote{ 
	In the European Union, dominating positions in telecommunications 
	markets (such as an ISP imposing side-payments to CPs at a price 
	of his choice) are controlled by the article 14, paragraph 3 of 
	the Directive 2009/140/EC, considering the application of remedies 
	to prevent the leverage of a large market power over a secondary 
	market closely related.} 
side payments, in either direction, and demand-dependent advertising 
revenues affect equilibrium usage-based prices. We also address equilibria 
in Stackelberg leader-follower dynamics. 
 
The rest of the paper is organized as follows. In section \ref{basic-sec}, we 
describe a basic model and derive Nash equilibria for competitive and 
collaborative scenarios. We consider potentially non-neutral side-payments
in section \ref{side-sec} and add advertising revenues in section \ref{ad-sec},
analyzing in each case how they impact equilibrium utilities.
{We study an ISP offering multiple service classes in section
\ref{diffserv-sec}, and generalize our model in section
\ref{multicp-sec} to non-monopolistic content or access providers.}
In section \ref{Stackelberg-sec}, we consider leader-follower dynamics.
We conclude in section \ref{conclusion} and discuss future work.

\section{Basic model} 
\label{basic-sec} 
 
Our model encompasses three actors: 
\begin{itemize} 
\item the internauts (users), collectively, 
\item a network access provider for the internauts, collectively 
called ISP1, and 
\item a content provider and its ISP, collectively called CP2. 
\end{itemize} 
The two providers play a game to settle on 
their (usage-based) prices. The internauts are modeled through 
their demand response. 
 
Consumers are assumed willing to pay a usage-based fee (which 
can be \$0/byte) for service/content that requires both providers. 
 
Denote by $p_i \geq 0$ the usage-based price leveed by 
provider $i$ (ISP1 being $i=1$ and CP2 being $i=2$). We assume that 
the demand-response of customers, which corresponds to the amount 
(in bytes) of content/bandwidth they are ready to consume 
given prices $p_1$ and $p_2$, follows a simple linear model: 
\begin{equation} 
D = D_0 - d (p_1 + p_2). 
\end{equation} 
With such a profile, we are dealing with a set of homogeneous 
users sharing the same response coefficient $d$ to price 
variations. Parameter $D_0$ corresponds to demand for zero
usage-based prices, which can be considered the demand under
pure \emph{flat-rate} pricing assuming that the usage-based
prices are overages on flat montly fees.

Demand should be non-negative, \ie 
\begin{equation*} 
p_1 + p_2\ \leq\ \frac{D_0}d\ =:\ \pmax. 
\end{equation*} 
Provider $i$'s usage-based revenue is given by 
\begin{equation} 
\label{ui-basic} 
U_i\ =\ D p_i. 
\end{equation}

\subsection{Competition} 
 
Suppose the providers do not cooperate. A Nash Equilibrium Point (NEP) 
$(p_1^*, p_2^*)$ of this two-player game satisfies: 
\begin{equation*} 
\frac{\partial U_i}{\partial p_i}(p_1^*, p_2^*) =  D^* - p_i^* d  = 0 
\quad \textrm{for } i=1,2, 
\end{equation*} 
which leads to $p_1^* = p_2^* = D_0 / (3d)$. 
The demand at equilibrium is thus $ D^* = D_0 / 3 $ 
and the revenue of each provider is 
\begin{equation} 
\label{ui-basic-comp} 
U_i^* = \frac{D_0^2}{9d}. 
\end{equation}

\subsection{Collaboration} 
 
Now suppose there is a coalition between ISP1 and CP2. 
Their overall utility is then $\Utot := U_1 + U_2 = Dp$, 
and an NEP $(p_1^*, p_2^*)$ satisfies 
\begin{equation*} 
\pd{\Utot}{p_i}(p_1^*, p_2^*)\ =\ D^* - d (p_1^* + p_2^*)\ =\ 0 
\quad \textrm{for } i=1,2, 
\end{equation*} 
which yields $p^* := p_1^*+p_2^* = D_0 / (2d)$. 
The demand at equilibrium is then $ D^* = D_0 / 2$, 
greater than in the non-cooperative setting. 
The overall utility $\Utot^* = D_0^2 / (4d)$ is also greater 
than $D_0^2 / (4.5 d)$ for the competitive case. 
Assuming both players share this revenue equally 
(trivially, the Shapley values are $\{1/2, 1/2\}$ in this case), 
the utility per player becomes 
\begin{equation} 
\label{ui-basic-collab} 
U_i^* = \frac{D_0^2}{8d}, 
\end{equation} 
which is greater than in the competitive case. 
So, both players benefit from this coalition.

\section{Side-Payments under Competition} 
\label{side-sec} 
 
Let us suppose now that there are \emph{side payments} 
between ISP1 and CP2 at (usage-based) price $p_s$. 
The revenues of the providers become: 
\begin{eqnarray} 
U_1 & = & D \left(p_1 + p_s\right) \\ 
U_2 & = & D \left(p_2 - p_s\right) 
\end{eqnarray} 
Note that $p_s$ can be positive (ISP1 charges CP2 for ``transit'' 
costs) or negative (CP2 charges ISP1, \eg for copyright 
remuneration\footnote{In France, a new law has been proposed recently to allow 
	download of unauthorized copyright content, and in return be charged 
	\emph{proportionally} to the volume of the download.}). 
It is expected that $p_s$ is \emph{not} a decision variable of 
the players, since their utilities are monotonic in 
$p_s$  and the player without control would 
likely set (usage-priced) demand to zero to avoid negative utility. 
That is,  
$p_s$ would normally be \emph{regulated} 
and we will consider it as a fixed parameter in the following  
(with $|p_s| \leq \pmax$). 
 
First, if $|p_s| \leq \frac13 \pmax$, the equilibrium prices are given by 
\begin{eqnarray*} 
	p_1^* &=& \frac13 \pmax - p_s \\ 
	p_2^* &=& \frac13 \pmax + p_s 
\end{eqnarray*} 
but demand $D^* = D_0 / 3$ and utilities 
\begin{equation*} 
U_i^* = \frac{D_0^2}{9d} 
\end{equation*} 
are exactly the same as \pref{ui-basic-comp} in the competitive setting with no side 
payment. Therefore, though setting $p_s>0$ at first seems to favor ISP1 
over CP2, it turns out to have no effect on equilibrium revenues for 
both providers. 
 
Alternatively,  
if $p_s \geq \frac13 \pmax$,  
a boundary Nash equilibrium is reached when $p_1^*=0$ 
and $p_2^*=\frac12 (\pmax + p_s)$, which means ISP1 does not charge 
usage-based fees to its consumers. Demand becomes 
$D^* = \frac12 (D_0 - d p_s)$, and utilities are 
\begin{eqnarray*} 
U_1^* &=& \frac{(D_0 - d p_s)d p_s}{2d}\\ 
U_2^* &=& \frac{(D_0 - d p_s)^2}{4d}  
\end{eqnarray*} 
Though $p_1^*=0$, $U_1^*$ is still strictly positive, with revenues for 
ISP1 coming from side-payments (and possibly from flat-rate monthly fees as well). 
Furthermore, $p_s \geq \frac13 \pmax \Leftrightarrow d p_s \geq \frac12 (D_0 - d p_s)$, 
which means $U_1^* \geq U_2^*$: in this setting, ISP1's best move is to set 
his usage-based price to zero (to increase demand), while he is sure to achieve 
better revenue than CP2 through side-payments. 
 
Finally, if $p_s < -\frac13 \pmax$, the situation is  
similar to the previous case (with $-p_s$ instead of $p_s$). 
So, here $p_2^*=0$ and $p_1^*=\frac12 (\pmax - p_s)$, 
leading to $U_2^* \geq U_1^*$. 
 
To remind, herein revenues $U_i$ are assumed usage-based, 
which means there could also be flat-rate charges in play to generate 
revenue for either party. Studies of flat-rate compare to 
usage-based pricing schemes  
can be found in the literature, see, \eg \cite{ciss08}.

\section{Advertising revenues} 
\label{ad-sec} 
 
We suppose now that CP2 has an additional source of (usage-based) revenue 
from advertising that amounts to $D p_a$. Here $p_a$ is not a decision 
variable but a fixed parameter.\footnote{ 
	One may see $p_a$ as the result of an independent game between 
	CP2 and his advertising sources, the details of which are out of  
	the scope of this paper.}

\subsection{Competition} 
 
The utilities for ISP1 and CP2 are now: 
\begin{eqnarray} 
\label{u1-complete} 
U_1 &=& \left[D_{0} - d \cdot \left(p_1+p_2\right)\right] \left(p_1+p_s \right) \\ 
\label{u2-complete} 
U_2 &=& \left[D_0 - d \cdot (p_1 + p_2)\right] (p_2 - p_s + p_a) 
\end{eqnarray} 
Here, the Nash equilibrium prices are: 
\begin{eqnarray*} 
p_1^* &=& \frac13 \pmax - p_s + \frac13 p_a \\ 
p_2^* &=& \frac13 \pmax + p_s - \frac23 p_a 
\end{eqnarray*} 
The cost to users is thus $p^* = \frac23 \pmax - \frac13 p_a$ while demand is 
$D^* = \frac13 (D_0 + d p_a)$. Nash equilibrium utilities are given by 
\begin{equation} 
U_i^*\ =\ \frac{(D_0 + d p_a)^2}{9d} \quad \textrm{for }i=1,2, 
\end{equation} 
which generalizes equation \pref{ui-basic-comp} and shows how advertising 
revenue quadratically raises players' utilities.

\subsection{Collaboration} 
 
The overall income for cooperating providers is 
\begin{equation} 
\Utot = (D_0 - dp)(p + p_a). 
\end{equation} 
So, solving the associated NEP equation yields 
\begin{equation} 
p^* = \frac{\pmax - p_a}{2}. 
\end{equation} 
The NEP demand is then $D^* = (D_0 + d p_a)/2 $, and the total revenue at 
Nash equilibrium is $\Utot^*=(D_0 + d p_a)^2/(4d)$. Assuming this revenue 
is split equally between the two providers, we get for each provider the equilibrium 
utility 
\begin{equation} 
U_i^*=\frac{(D_0 + d p_a)^2}{8d}, 
\end{equation} 
which generalizes equation \pref{ui-basic-collab}. As before, providers and users are 
better off when they cooperate. 
 
Thus, we see that $p_a>0$ leads to lower prices, increased demand and more 
revenue for \emph{both} providers (\ie including ISP1).

\section{ISP Providing Multiple Service Classes}
\label{diffserv-sec}

In this section, we suppose ISP1 is offering two types of network access service: a low-quality one $l$ at price $p_l$, and a high-quality one $h$ at price $p_h \geq p_l$. The role of multiple service classes in a neutral network has previously 
been explored, \eg in \cite{NOMS10}. Here, we split the demand $D$ into $D_l$ and $D_h$: $D = D_l + D_h$ (we will describe later how we implement the dichotomy between $D_l$ and $D_h$). For now, assume the overall demand still has a linear response profile, \ie
\begin{equation}
D = D_0 - d (\underbrace{p_l + p_h}_{\textrm{formerly } p_1} + p_2).
\end{equation}

First, we make reasonable assumptions on $D_l$:
\begin{enumerate}
\item \emph{Pricing incentives:} Define $\Delta p := p_h - p_l$. $\Delta p$ is an incentive for consumers to chose between classes $l$ and $h$: 
the higher $\Delta p$ is, the more likely users are to select $l$. Thus, if we take $x := 1/{\Delta p}$ and $y := D_l / D$, we may see $y$ as a function of $x$ and model this pricing response with the following properties:
\begin{eqnarray}
\label{cond-incr}
y'(x) & \leq & 0 \quad(D_l \textrm{ increases with } \Delta p)\\
\label{cond-zero}
y(0) &=& 1 \quad(D_l \uparrow D \textrm{ as } \Delta p \uparrow \infty) \\
\label{cond-infty}
y(\infty) &=& 0 \quad(D_l \downarrow 0 \textrm{ as } \Delta p \downarrow 0)
\end{eqnarray}

\item \emph{Congestion incentives:} As $D_l$ approaches $D$, we assume congestion occurs in the low-quality network, further deterring users to chose it. This motivates the additional assumption that
\begin{equation}
\label{cond-cgst}
|y'(x)| \downarrow 0 \textrm{ as } x \downarrow 0,
\end{equation}
that is, $D_l$ \emph{decelerates} as it gets closer to $D$.

\end{enumerate}
Define
\begin{equation}
\delta\ :=\ \frac{\Delta p}{\gamma \pmax},
\end{equation}
where $\gamma>0$ is an additional users' price-sensitivity parameter. The following 
demand relation satisfies all conditions \pref{cond-incr}, \pref{cond-zero}, \pref{cond-infty} and \pref{cond-cgst}:
\begin{equation}
D_l := \left(1 - e^{-\delta}\right) D.
\end{equation}
The providers' utilities are then:
\begin{eqnarray}
U_1 & = & D_l p_l + D_h p_h = D \left(p_l + \Delta p e^{-\delta}\right) \\
U_2 & = & D p_2
\end{eqnarray}

\subsection{Collaboration}\label{multiple-collab-subsec}

If both players cooperate, their overall utility is
\begin{eqnarray*}
\Utot & = & D \left(p_2 + p_l + \Delta p e^{-\delta}\right).
\end{eqnarray*}
There is no NEP with strictly positive prices 
$p_i \geq 0$ for $i=1,2$. To specify the boundary NEP  
(where at least one usage-based price is zero),
define
\[
\phi(x) \ := \ (1-x) e^{-x}
\]
and note that $\phi$ is a bijection of $[0,1]$.
\begin{itemize}
\item If $p_2=0$, NEP conditions imply
	\begin{eqnarray*}
	\delta^* &=& \phi^{-1}(1/2) \\
	p_l^* & = & \frac13 \left(\frac12 - \gamma \delta e^{-\delta}\right) \pmax
	\end{eqnarray*}
Utility at the NEP is therefore
	\begin{equation}
	\Utot^* \ = \ \frac{D_0^2}{9d} \left[\frac12 + 2 \gamma \delta e^{-\delta}\right].
	\left[2 + \left(2 e^{-\delta} - 3\right) \delta \gamma\right]
	\end{equation}
In this setting, the value of $\Utot$ is upper bounded by
$\approx~0.162 \frac{D_0^2}{d}$ which is achieved when
$\gamma \approx 1.53$ (recall that $\gamma$ is not
a decision variable).
\item If $p_l=0$, then $p_h=0$ and $p_2=\frac12\pmax$,
yielding
	\begin{equation}
	\Utot \ = \ \frac{D_0^2}{4d}.
	\end{equation}
\end{itemize}
Hence, irrespective of consumers' sensitivity $\gamma$ to the price gap $\Delta p$, the best solution for the coalition is to set-up usage-based pricing for content only, at price $p_2=\pmax / 2$, while network access is subject only to 
flat-rate pricing ($p_l=p_h=0$).

\subsection{Splitting Demand-Response Coefficient}

\begin{figure}[!ht]
	\includegraphics[width=0.5\columnwidth]{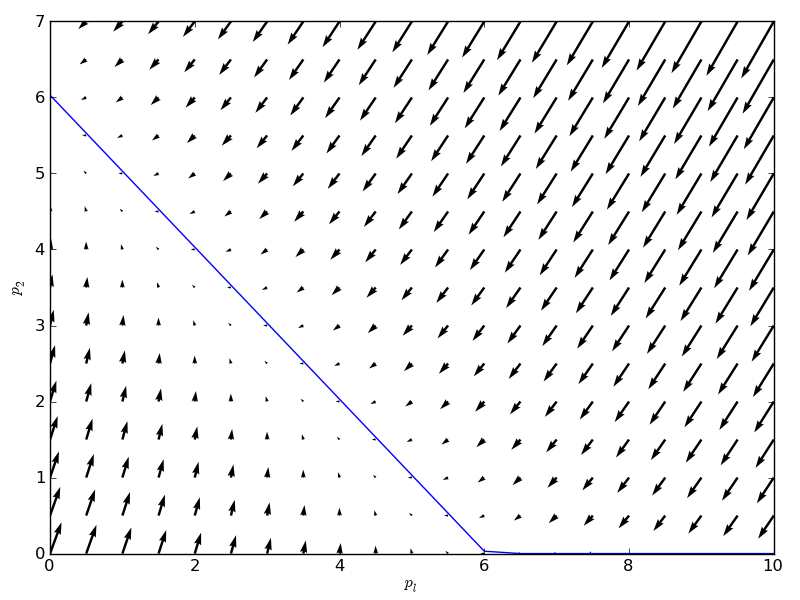}
	\caption{Attraction of the equilibrium line.}
	\label{fig-split}
	\end{figure}

Now consider splitting the demand-response coefficient 
$d$ into $d_l$, $d_h$ and $d_2$, that is:
\begin{equation}
D = D_0 - d_l p_l - d_h p_h - d_2 p_2.
\end{equation}
If
\begin{eqnarray}
\label{response-rel} d_2 &=& d_l + d_h ,
\end{eqnarray}
then  the interior equilibrium conditions $\nabla \Utot = \vec{0}$ yield:
\begin{eqnarray*}
\delta &=& \phi^{-1}(d_h / d_2) \\
p_l + p_2 &=& \frac{D_0}{2 d_2} - \frac{\delta {\Delta p}_0}2 \left(\frac{d_h}{d_2} + e^{-\delta}\right)
\end{eqnarray*}
When the demand-response coefficients satisfy \pref{response-rel}, 
we have an \emph{equilibrium line}.
Vector field plots of $\Utot$ suggest it is 
attractive (see Figure \ref{fig-split}). 
In this particular setting, 
providers can thus reach $\Utot^*$ with usage-based pricing.

However, if $d_2 \neq d_l + d_h$, there exists a line of attraction, 
but with a non-null gradient on it driving players toward border 
equilibria. 
Hence, the conclusion of subsection \ref{multiple-collab-subsec} 
also holds in this more generalized setting.

\subsection{Competition}

\begin{figure}[h!]
	\includegraphics[angle=-90,width=0.5\columnwidth]{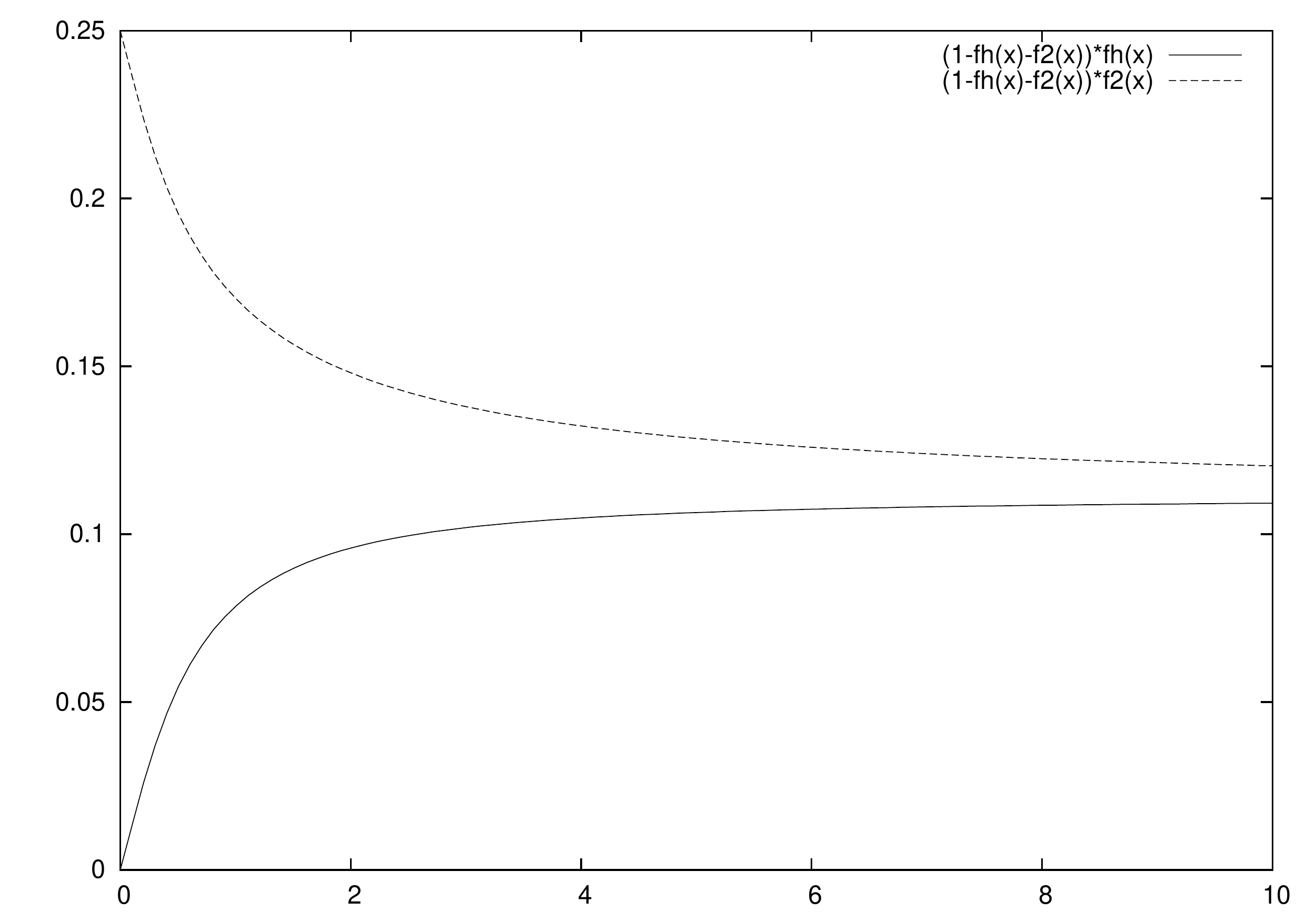}
	\caption{Utilities as functions of users' sensitivity to usage-based pricing.}
	\label{fig-ut-1}
	\end{figure}

When ISP1 and CP2 compete, again there is no interior NEP (with 
all prices $p_i$ strictly positive).
In fact, the condition
$\nabla_{p_l,p_h} U_1 = \vec{0}$ implies $p_l=0=p_h$ and $D=0$, so ISP1
has to relax condition $\pd{U_1}{p_l}=0$ by setting $p_l=0$ 
(\ie only flat-rate pricing for the best-effort service $l$). The solution to the two remaining Nash
equilibrium conditions is then:
\begin{eqnarray}
p_2 &=& \frac14 \left[\sqrt{9 \gamma^2 + 2\gamma + 1} - 3\gamma + 1\right] \cdot \pmax \\
p_h &=& \frac\gamma{2 \sqrt{9\gamma^2 + 2\gamma + 1} - 3\gamma + 2} \cdot \pmax
\end{eqnarray}
By defining $f_2(\gamma) := p_2 / \pmax$ and $f_h(\gamma) := p_h / \pmax$, 
we then have
	\begin{eqnarray*}
	U_1^*(\gamma) &=& f_h(\gamma) \cdot (1 - f_h(\gamma) - f_2(\gamma)) D_0 \pmax \\
	U_2^*(\gamma) &=& f_2(\gamma) \cdot (1 - f_h(\gamma) - f_2(\gamma)) D_0 \pmax
	\end{eqnarray*}
Figure \ref{fig-ut-1} shows utilities at equilibrium (as fractions of $D_0 \pmax$). We see that, in any case, CP2 has
the advantage in this game: $U_2^*$ is always greater to 
$U_1^*$, irrespective of consumers' sensitivity $\gamma$ 
to usage-based prices.

Here, $\gamma \to 0$ means users are so sensitive to any usage-based price that they will always choose the best-effort service 
(which is subject to flat-rate pricing).  
Users' price sensitivity decreases as $\gamma$ increases, the limit 
$\gamma \to \infty$ corresponding to the setting of section \ref{basic-sec} 
with $\lim_{\gamma\rightarrow\infty} U_i^*(\gamma) 
=
\frac{D_0^2}{9d}$.

\section{Multiple CPs providing the same type of content}
\label{multicp-sec}

Now suppose there are multiple CPs supplying the \emph{same} type of content (\eg competing online encyclopedias), so users choose one CP over another based only on price.

For the sake of simplicity, let us consider the case with two CPs denoted by CP2 and CP3. First, let us remark that if there is a significant difference between the prices $p_2$ of CP2 and $p_3$ of CP3, since both provide the same type of content, all consumers are likely to shift to the cheapest provider, leading us back to our initial model with one ISP and one CP.

So the difference introduced by multiple CPs may arise when $p_2 \approx p_3$. Suppose that, initially,
	\begin{equation*}
	p_2=\bp=p_3.
	\end{equation*}
In this case, we assume customers are evenly shared between CP2 and CP3, so that
	\begin{equation*}
	U_i=\frac12 D(p_1,\bp) \bp \quad \textrm{for } i=2,3,
	\end{equation*}
where $D(p_1,p_i) := D_0 - d(p_1 + p_i)$ is the demand-reponse to the usage-based prices $p_1$ (for network access) and $p_i$ (for content).

Now, if CP$i$ reduces its price by some small $\delta p_i$, some of its opponent's consumers will change CP, but not all of them since a small price gap may not convince them to go. This behavior is known as \emph{customer stickiness, inertia or loyalty}. To model it we rewrite $U_i$ as
	\begin{equation}
	\label{sticky-ui}
	U_i \ = \ s(p_i, p_{5-i}) D(p_1, p_i) p_i	\quad \textrm{for } i=2,3,
	\end{equation}
where $p_{5-i}$ denotes the usage-based price of the other CP, and the ``stickiness'' function $s$ has the following properties:
	\begin{eqnarray}
	\label{sticky-s1} s(x, y) & \geq & 0, \\
	\label{sticky-s2} s(x, x) & = & \frac12, \\
	\label{sticky-s3} s(x, y) + s(y, x) & = & 1.
	\end{eqnarray}
When CP$i$ reduces its price by $\delta p_i$, the first-order variation in its utility is given by $\pd{U_i}{p_i}(\bp, \bp) \delta p_i$. From \pref{sticky-ui} and \pref{sticky-s3},
	\begin{equation*}
	\pd{U_i}{p_i} (\bp, \bp) = \left[
		\pd{s}{x}(\bp, \bp) +  \frac{\pmax - p_1 - 2 \bp}{2\bp (\pmax - p_1 - \bp)}
		\right] D(p_1, \bp) \bp.
	\end{equation*}
(Where $\pmax$ was defined as $\frac{D_0}{d}$.) Thus, taking consumers loyalty into consideration, the Nash equilibrium condition for either CP$i$ becomes:
	\begin{equation}
	\label{sticky-equ}
	\pd{s}{x}(\bp, \bp) \ + \ \frac{\pmax - p_1 - 2 \bp}{2\bp (\pmax - p_1 - \bp)} = 0.
	\end{equation}

\subsection{Stickiness Model 1}
\label{s1-subsec}

As a first, simple loyalty model, suppose that after CP$i$ reduces its price by $\delta p_i$, the fraction of users that remain with the other CP$(5-i)$ is inversely proportional to its price $p_{5-i}$, \ie the stickiness function is
	\begin{equation}
	\label{sm1}
	s(p_i, p_{5-i}) \ := \ \frac{1/p_i}{1/p_i + 1/p_{5-i}} \ = \ \frac{p_{5-i}}{p_i + p_{5-i}}.
	\end{equation}
In this setting, equilibrium condition \pref{sticky-equ} becomes $\bp = \frac13 (\pmax - p_1)$, while equilibrium condition for ISP1 is $p_1 = \frac12 (\pmax - \bp)$. Thus, prices at the NEP are $p_1^* = \frac25 \pmax$ and $\bp^* = \frac15 \pmax$. Demand at equilibrium is $D^* = \frac{2 D_0}{5d}$ and the revenue of each provider is
	\begin{eqnarray}
	\label{sm1-u1} U_1^* & = & \frac{4}{25} \Umax, \\ 
	\label{sm1-ui} U_i^* & = & \frac{1}{25} \Umax \quad \textrm{for } i=2,3. 
	\end{eqnarray}
We see that, compared to \pref{ui-basic-comp}, ISP1 highly benefits from the competition between CPs (his revenue is about $44\%$ higher). The situation would be symmetric with a single CP and two competing ISPs.

\subsection{Stickiness Model 2}

One can consider another loyalty model where the fraction of users remaining with CP$-i$ is proportional to the price \emph{slackness} $\pmax - p_{5-i}$, \ie the stickiness function is
	\begin{equation*}
	s(p_i, p_{5-i}) \ := \ \frac{\pmax - p_i}{(\pmax - p_i) + (\pmax - p_{5-i})}
	\end{equation*}
Here condition \pref{sticky-equ} becomes $2 (\pmax - \bp) (\pmax - p_1 - 2\bp) = \bp (\pmax - p_1 - \bp)$, while equilibrium condition for ISP1 is still $2 p_1 + \bp = \pmax$. Resolution of this system leads to $p_1^* = \frac{5}{14} \pmax$ and $\bp^* = \frac27 \pmax$. At the NEP, demand is thus $D^* = \frac{5}{14} D_0$ and utilities are
	\begin{eqnarray*}
	U_1^* & = & \frac{25}{196} \, \Umax \ \approx \ 0.12 \, \Umax, \\
	U_i^* & = & \frac{10}{196} \, \Umax \ \approx \ 0.051 \, \Umax \quad \textrm{for } i=2,3.
	\end{eqnarray*}
We see with this second setting that the outcome of the price war between CPs and ISP1 significantly depends on the customer inertia model used.

\subsection{Stickiness with side-payments}

Now focusing on the first stickiness model \pref{sm1} and update the model to take into account usage-based side payments $p_s$. The revenues become
	\begin{eqnarray*}
	U_1 & = & D (p_1 + p_s), \\
	U_i & = & s(p_i, p_{5-i}) D (p_i - p_s) \quad \textrm{for } i=2,3.
	\end{eqnarray*}
For same-priced CPs, solving $\pd{U_i}{p_i}(\bp, \bp) = 0$ (with non-nul demand), we find that the equilibrium conditions are
	\begin{eqnarray*}
	(\bp - p_s)(\pmax - p_1 + \bp) & = & 2 \bp (\pmax - p_1 - \bp), \\
	2 p_1 & = & \pmax - \bp - p_s.
	\end{eqnarray*}
They are now quadratic in $\bp$, thus, for the sake of readability, let us define
	\begin{eqnarray*}
	\eta & := & p_s / \pmax, \textrm{ and} \\
	\psi(\eta) & := & \sqrt{1 + 28 \eta + 36 \eta^2}.
	\end{eqnarray*}

\begin{figure}[ht]
	\includegraphics[width=0.5\columnwidth]{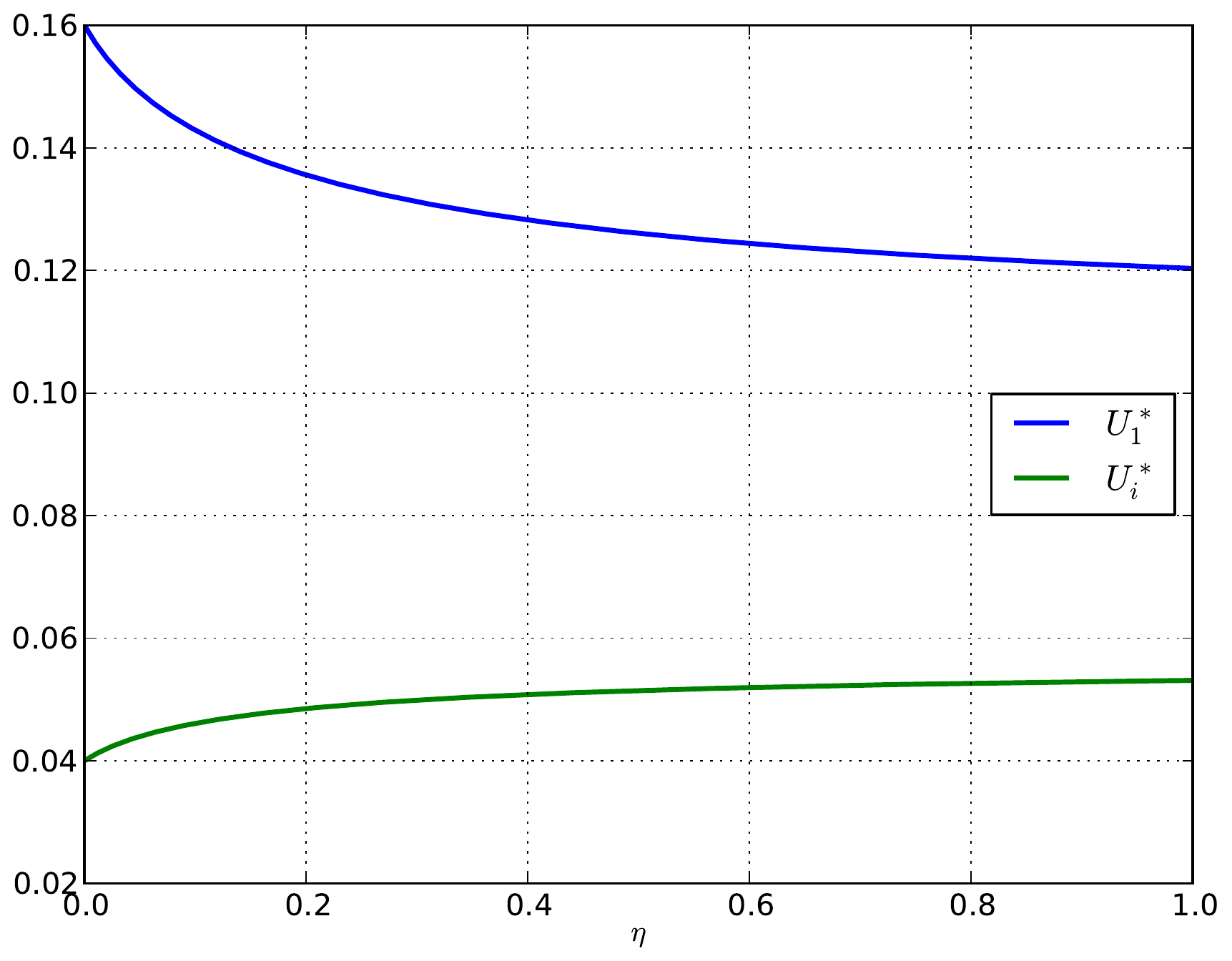}
	\caption{Revenues at the NEP as functions of $\eta := p_s / \pmax$.}
	\label{u-eta-1}
	\end{figure}

When $p_s>0$ (side payments from the CPs to the ISP), resolution of this system for positive prices lead us to:
	\begin{eqnarray*}
	\bp^* & = & \frac{\pmax}{10} \left(1 + 4 \eta + \psi(\eta) \right) \\
	p_1^* & = & \frac{\pmax}{20} \left(9 - 14 \eta - \psi(\eta) \right)
	\end{eqnarray*}
Then, demand at the NEP is $D^* = \frac{D_0}{20} (9 + 6 \eta - \psi(\eta))$ while revenues are
	\begin{eqnarray}
	U_1^* & = & \frac{\Umax}{400} (9 + 6 \eta - \psi(\eta))^2 \\
	U_i^* & = & \frac{\Umax}{100} (2 - 19 \eta - 18 \eta^2 + (2 + 3 \eta)\psi(\eta))
	\end{eqnarray}
What is interesting here is that both utilities are monotone in $\eta$ (see Figure \ref{u-eta-1}): $U_1^*$ decreases while $U_i^*$ increases with $\eta$ (when $p_s \ll \pmax$, we fall back to the results \pref{sm1-u1} and \pref{sm1-ui} of subsection \ref{s1-subsec}).
Paradoxically enough, we see that increasing $p_s$ (which means more usage-based side payments for ISP1) is disadvantageous for ISP1 but benefits the CPs! This situation is very different from the one in section \ref{side-sec}, where the ISP was favored over the CP when $\eta$ was over a fixed threshold.

When $p_s<0$, CPs receive usage-based side payments from ISP1 (ostensibly for royalties of copyrighted content).

If $p_s \leq \frac{-7 + 2 \sqrt{10}}{18} \pmax$, then $\pd{U_i}{p_i}(\bp,\bp)$ is always negative and $\bp$ will tend to zero. This means that the best strategy for CPs is to offer their content only for a flat rate, thus increasing demand and making all their usage-based profits on side payments.

Otherwise, if $p_s > \frac{-7 + 2 \sqrt{10}}{18} \pmax$, then condition \pref{sticky-equ} has two solutions:
	\begin{eqnarray*}
	\bp_0 & = & \frac{1}{10} (4 \eta + 1 - \psi(\eta)) \pmax, \\
	\bp_1 & = & \frac{1}{10} (4 \eta + 1 + \psi(\eta)) \pmax.
	\end{eqnarray*}
There are therefore two equilibria:
\begin{itemize}

\item $\bp^* = \bp_1$ and $p_1^* = \frac{1}{20} (-14 \eta + 9 - \psi(\eta))$: this is the case we studied in the $p_s>0$ setting, demand and revenues at the NEP are unchanged. This equilibrium is ``stable'' in the sense that\footnote{Recall that we restrict our attention to $p_2 \approx p_3$.} $\pd{U_i}{p_i}(\bp_1-, \bp_1-) > 0$ and $\pd{U_i}{p_i}(\bp_1+, \bp_1+) < 0$ for $i \in \{2,3\}$: if CPs move slightly their prices around $\bp^*=\bp_1$, they are incented to move back.

\item $\bp^* = \bp_0$ and $p_1^* = \frac{1}{20} (-14 \eta + 9 + \psi(\eta))$: in this case, demand at equilibrium is $D^* = \frac{D_0}{20} (9 + 6 \eta + \psi(\eta))$ and revenues are given by
	\begin{eqnarray}
	U_1^* & = & \frac{\Umax}{400} (9 + 6 \eta + \psi(\eta))^2, \\
	U_i^* & = & \frac{\Umax}{100} (2 - 19 \eta - 18 \eta^2 - (2 + 3 \eta)\psi(\eta)).
	\end{eqnarray}
However, this equilibrium is ``unstable'' in the sense that $\pd{U_i}{p_i}(\bp_1-, \bp_1-) < 0$ and $\pd{U_i}{p_i}(\bp_1+, \bp_1+) > 0$: if CPs shift their prices from $\bp^*=\bp_0$, they are incented to shift even more, which will lead them either to the other equilibrium or to $\bp=0$ (again, no usage-based pricing for content, but there may additional revenue from flat-rate fees as well).
	\begin{figure}[ht]
	\includegraphics[width=0.5\columnwidth]{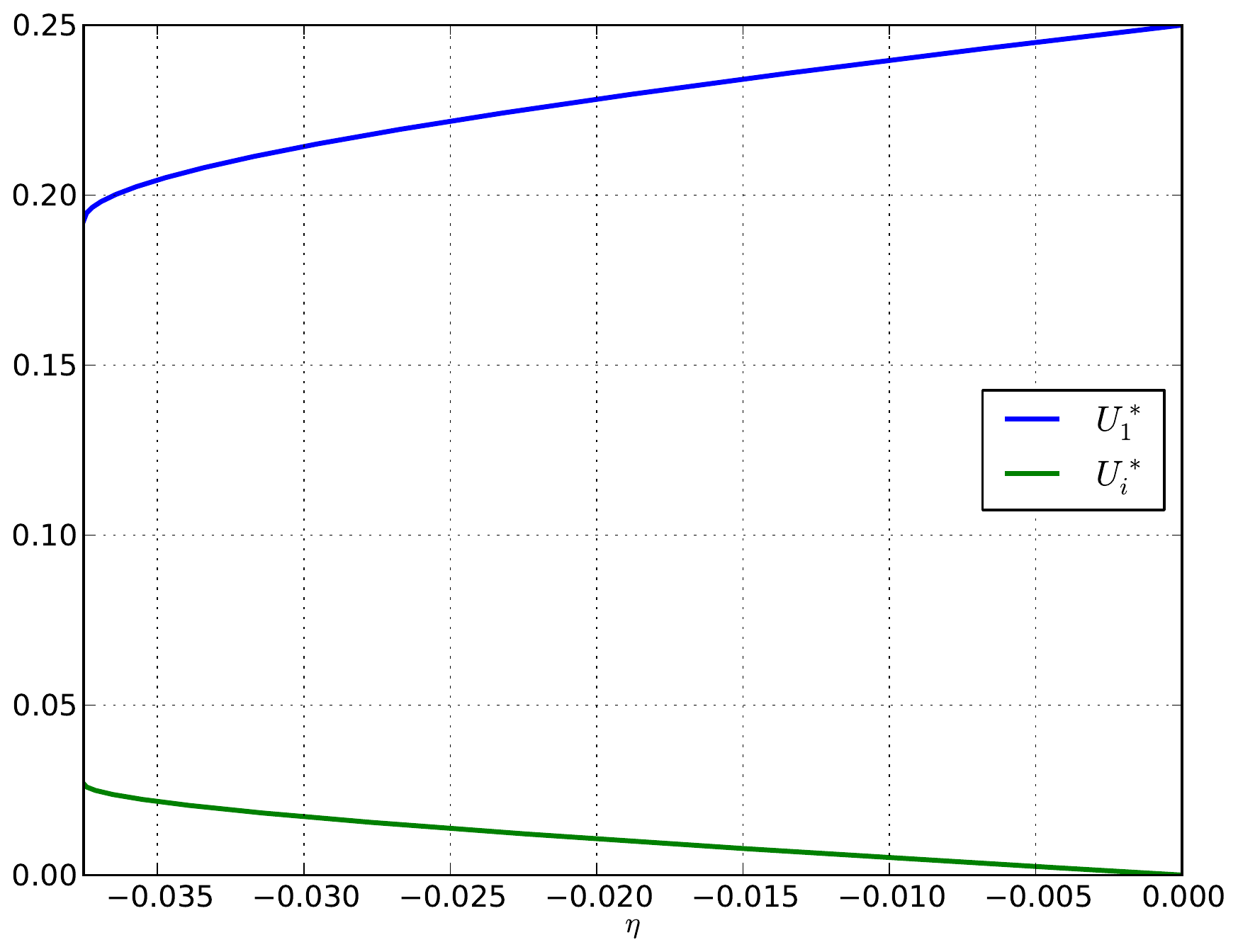}
	\caption{Revenues at the unstable NEP as functions of $\eta$.}
	\label{fig-u-eta-2}
	\end{figure}

\end{itemize}
The ISP is better off at this new NEP (see Figure \ref{fig-u-eta-2}): regarless of the (regulated) value of $p_s$, his revenue is always higher here (and the CPs' revenues are always lower) than at the other NEP. This fact is consistent with the ``unstability'' we observed: if the CPs happen to leave this equilibrium, they are not incented to come back.

A similar story follows if one considers multiple competing ISPs with one CP. Also, taking advertising revenues into consideration will complicate the above computations and affect the location and ``stability'' of the NEPs.

\section{Stackelberg equilibrium} 
\label{Stackelberg-sec} 
 
Stackelberg equilibrium corresponds to asymmetric competition in which one 
competitor is the leader and the other a follower. Actions are no longer taken 
independently: the leader takes action first, and then the follower reacts. 
 
Though the dynamics of the games are different from the previous study, 
equations \pref{u1-complete} and \pref{u2-complete} still hold, with fixed 
$p_a \geq 0$ and regulated $p_s$. In the following, we need to assume that 
	\begin{eqnarray*} 
	p_s & \leq & \frac12 \pmax + \frac12 p_a \\ 
	p_a & \leq & \frac13 \pmax + \frac14 p_s 
	\end{eqnarray*} 
so that NEPs are reachable with positive prices. 
 
If ISP1 sets $p_1$, then CP2's optimal move is to set 
	\begin{equation*} 
	p_2\ =\ \frac12 (-p_1 + \pmax + p_s - p_a). 
	\end{equation*} 
This expression yields $D = \frac{d}{2}(\pmax - p_1 - p_s + p_a)$ 
and $U_1 = \frac{d}{2}(\pmax - p_1 - p_s + p_a)(p_1 + p_s)$. 
Anticipating CP2's reaction in trying to optimize $U_1$, the best 
move for ISP1 is thus to set 
	\begin{equation*} 
	p_1^*\ =\ \frac{1}{2} \pmax -  p_s + \frac{1}{2} p_a, 
	\end{equation*} 
which yields 
	\begin{equation*} 
	p_2^* = \frac14 \pmax + p_s - \frac34 p_a. 
	\end{equation*} 
Therefore, when ISP1 is the leader, at the NEP demand is $D^* = \frac14 (D_0 + d p_a)$ 
and utilities are: 
	\begin{eqnarray} 
	U_1^* &=& \frac{1}{8d} (D_0 + d p_a)^2, \\ 
	U_2^* &=& \frac{1}{16d} (D_0 + d p_a)^2. 
	\end{eqnarray} 
Suppose now that CP2 is the leader and sets $p_2$ first. 
Similarly, we find: 
	\begin{eqnarray*} 
	p_2^* &=& \frac12 \pmax + p_s - \frac12 p_a \\ 
	p_1^* &=& \frac14 \pmax - p_s + \frac14 p_a 
	\end{eqnarray*} 
These values yield the same cost $p^*$ and demand $D^*$ 
for the internauts at the NEP, while providers' utilities become: 
	\begin{eqnarray} 
	U_1^* &=& \frac{1}{16d} (D_0 + d p_a)^2, \\ 
	U_2^* &=& \frac{1}{8d} (D_0 + d p_a)^2. 
	\end{eqnarray} 
 
Therefore, in either case of leader-follower dynamics, the leader 
obtains twice the utility of the follower at the NEP (yet, his revenue is 
not better than in the collaborative case).

\section{Conclusions and on-going work} 
\label{conclusion} 
 
Using a simple model of linearly diminishing consumer demand as a 
function of usage-based price, we studied a game between a monopolistic ISP 
and a CP under a variety of scenarios including consideration of: 
non-neutral two-sided transit pricing (either CP2 participating in network 
costs or ISP1 paying for copyright remuneration), 
advertising revenue, competition, cooperation and leadership. 
 
In a basic model without side-payments and advertising revenues, 
both providers achieve the same utility at equilibrium, and all actors 
are better off when they cooperate (higher demand and providers' 
utility). 
 
When regulated, usage-based side-payments  
$p_s$ come into play, the outcome 
depends on the value of $|p_s|$ compared to the maximum usage-based 
price $\pmax$ consumers can tolerate: 
\begin{itemize} 
\item when $|p_s| \leq \frac13 \pmax$, providers shift their prices to fall back to 
the demand of the competitive setting with no side-payments; 
\item when $|p_s| \geq \frac13 \pmax$, the provider receiving side payments 
sets its usage-based price to zero to increase demand, while it is sure 
to be better off than his opponent. 
\end{itemize} 
 
When advertising revenues to the CP come into play, they increase the 
utilities of \emph{both} providers by reducing the overall usage-based 
price applied to the users. ISP1 and CP2 still share the same utility at 
equilibrium, and the increase in revenue due to advertising is quadratic. 
 
{We considered in section \ref{diffserv-sec} the implications of 
\emph{service differentiation} from the ISP. In our model, when ISP1 and 
CP2 cooperate, the best solution for them is to set-up usage-based prices 
for content only and flat-rate pricing for network access. 
However, when providers do not cooperate, the ISP optimally offers its 
best-effort service for a flat rate (zero usage-based cost),  
resulting in more usage-based revenue for the CP.}

{We considered in section \ref{multicp-sec} a generalization of our model
to non-monopolistic, competing CPs. For a simple customer inertia model,
we found that regulated side-payments had a significant impact on equilibrium
revenues for the ISP and the CPs:
	\begin{itemize}
	\item when side payments go to the access provider,
	his utility at the NEP diminishes, while
	\item when they go to the content providers,
	the three-player system has two equilibria: an
	unstable one in favor of the ISP, and a stable one.
	\end{itemize}}
 
Under leader-follower dynamics, the leader obtains twice the utility of his 
follower at equilibrium; yet, he does not achieve a better revenue than 
in the cooperative scenario. 
 
{In on-going work, we are exploring the effects of 
content-specific (\ie not \emph{application} neutral) pricing, 
including multiple CPs providing different types of content.}

\end{document}